\documentclass[conference]{IEEEtran}
\IEEEoverridecommandlockouts

\usepackage{balance}
\usepackage{enumitem}
\usepackage{here}
\usepackage{bm}
\usepackage{booktabs}
\usepackage{tabularx}
\usepackage{subfigure}
\usepackage{caption}
\usepackage{subcaption}
\usepackage{cite}
\usepackage{amsmath,amssymb,amsfonts}
\usepackage{algorithmic}
\usepackage{graphicx}
\usepackage{textcomp}
\usepackage{comment}
\usepackage{xcolor}
\def\BibTeX{{\rm B\kern-.05em{\sc i\kern-.025em b}\kern-.08em
    T\kern-.1667em\lower.7ex\hbox{E}\kern-.125emX}}
\begin{document}

\title{Measurement-based Characterization of ISAC Channels with Distributed Beamforming at Dual mmWave Bands and with Human Body Scattering and Blockage \\
}

\author{\IEEEauthorblockN{Yang MIAO}
\IEEEauthorblockA{\textit{Faculty of Electrical Engineering} \\
\textit{University of Twente}\\
Enschede, the Netherlands \\
y.miao@utwente.nl}
\and
\IEEEauthorblockN{Minseok Kim, Naoya Suzuki}
\IEEEauthorblockA{\textit{School of Science and Technology} \\
\textit{Niigata University}\\
Niigata, Japan \\
mskim@eng.niigata-u.ac.jp}
\and
\IEEEauthorblockN{Chechia Kang, Junichi Takada}
\IEEEauthorblockA{\textit{Dept. of Transdisciplinary Science and Engineering} \\
\textit{Tokyo Institute of Technology}\\
Tokyo, Japan \\
kang.c.aa@m.titech.ac.jp}
}

\maketitle

\begin{abstract}
In this paper, we introduce our millimeter-wave (mmWave) radio channel measurement for integrated sensing and communication (ISAC) scenarios with distributed links at dual bands in an indoor cavity; we also characterize the channel in delay and azimuth-angular domains for the scenarios with the presence of 1 person with varying locations and facing orientations. In our setting of distributed links with two transmitters and two receivers where each transceiver operates at two bands, we can measure two links whose each transmitter faces to one receiver and thus capable of line-of-sight (LOS) communication; these two links have crossing Fresnel zones. We have another two links capable of capturing the reflectivity from the target presenting in the test area (as well as the background).  
The numerical results in this paper focus on analyzing the channel with the presence of one person. It is evident that not only the human location, but also the human facing orientation, shall be taken into account when modeling the ISAC channel.
\end{abstract}


\section{Introduction}
The 5G NR and IEEE 802.11ay standards have already included the mmWave bands, aiming to provide wireless communications with ultra-high speed and ultra-low delay to serve for demanding use cases like virtual reality, industrial 5.0 and autonomous driving. The 5G NR mmWave bands start from 24 GHz and go up to 71 GHz \cite{5GNRmmWave}. The WiGig targets the 60 GHz band starting from 58 GHz to 74 GHz \cite{WiGig}. These two standards are expected to co-exist in the complex and dynamic indoor environment, where the influences of human body and body dynamics are often unpredictable, making the indoor mmWave communications challenging in practice.
Moving forward to 6G, researchers have been focusing on ISAC, where both high-speed communication and accurate sensing of the surrounding environment are important for the envisioned use cases like remote health monitoring, autonomous driving, and industry 5.0. Dual-functionality including communication and sensing will be integrated into future radio systems \cite{CoDesign1, CoDesign2}. Sensing is especially promising at the mmWave bands of the 5G NR and WiGig standards, where the smaller wavelength, the larger bandwidth and the array size/complexity are converging to radars that are conventionally used for sensing.

The environment seen by 5G NR and WiGig systems is however much more dynamic and complex than the environment seen by outdoor radars. To enable ISAC in mmWave 5G NR and WiGig applications in dynamic indoor environments, an accurate modeling of human body interactions with the mmWaves is crucial. Radio waves are potentially reverberating \cite{reverberation} between multiple indoor reflectors or human; such interactions, given a multi-link scenario, involve not only absorption but also scattering (bistatic scattering or backscattering).  
In particular, one of the ISAC use cases aims to prevent mmWave communications from human blockage, hence the proactive sensing (detection, localization and tracking) of humans before any blockage is the goal, wherein the human body serves as a scatterer/reflector in radar sensing. However, relevant works have mainly focused on the modeling of human blockage \cite{Blockage4} and skin reflectivity \cite{SKINReflectance}; multi-link and multi-band characterizations in mmWave ISAC scenarios with the presence of humans are largely unexplored.

In this paper we present our dual-band and distributed MIMO beamforming measurement as well as the characterization of the channels with the presence of one person with controlled location and facing orientations. Novelties are:
\begin{enumerate}
\item Our novel measurement setup includes (i) a dual-band mmWave channel sounder which can measure at 24 GHz and 60 GHz concurrently with a $2\times2$ distributed MIMO topology, and (ii) a distributed RGB-Depth (RGBD) camera system which captures the depth image of groundtruth via a global registration procedure to obtain point clouds from distributed local to a global coordinate.
\item Our novel indoor measurement scenarios have the presence of up-to-3 persons, located at eight different positions in relative to the arrays, separated by different distances, and facing eight directions (body orientation).
\item We present the characterization results of the distributed dual-band channels with the presence of 1 person, in both the delay and the azimuth-angular domains, and discuss our observations and insights for ISAC channel modeling. 
\end{enumerate}

\section{Dual-Band Distributed-Beamforming Indoor Channel Measurement for ISAC Scenarios}

\subsection{The Dual-Band Distributed MIMO Channel Sounder}

\begin{figure}[tb]
\centering
\includegraphics[width=0.8\columnwidth]{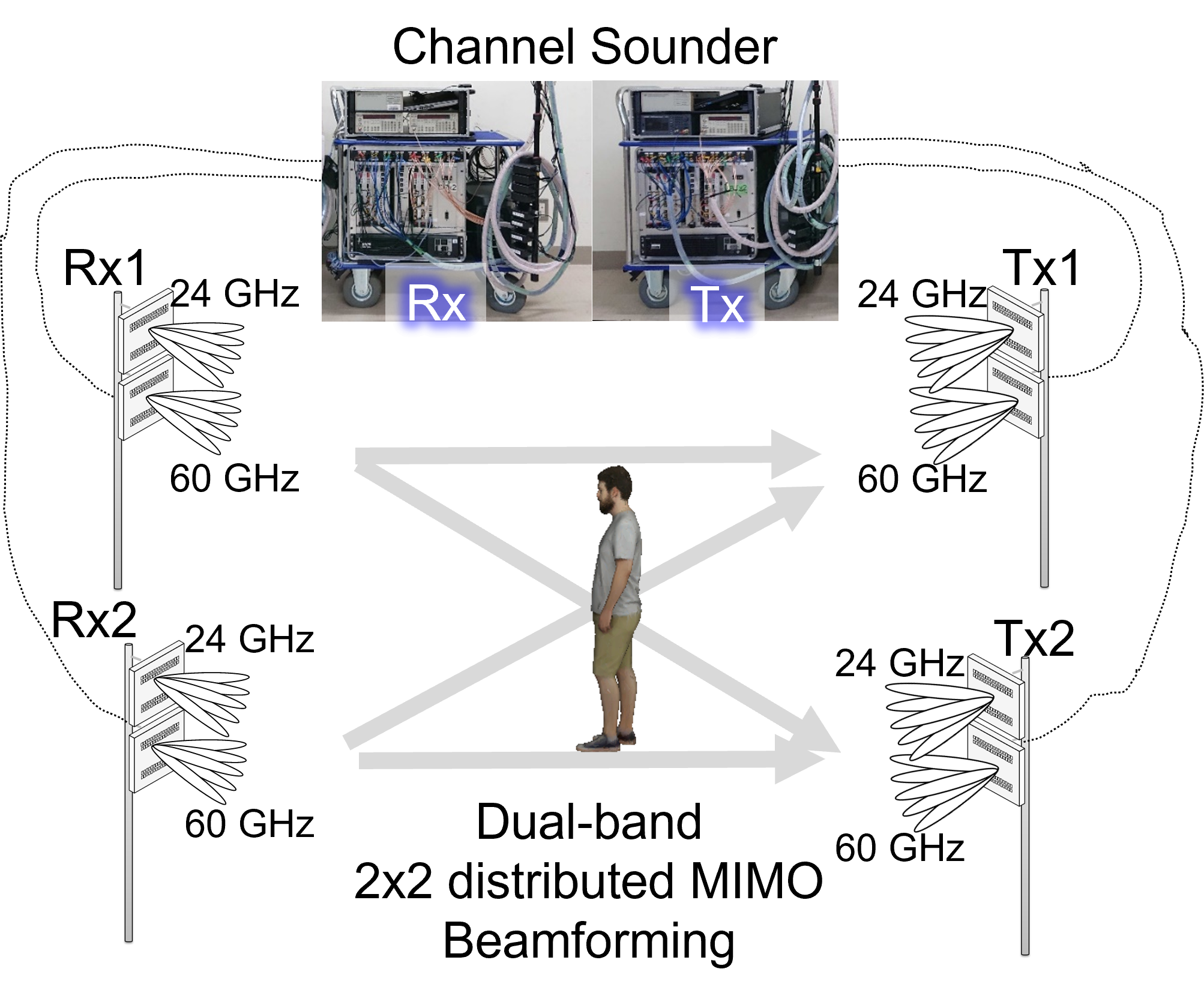}
\caption{System configuration of dual-band distributed MIMO channel sounder}
\label{fig:sounderimage}
\end{figure}

Fig.~\ref{fig:sounderimage} shows the channel sounder developed at the Niigata University \cite{Sounder}, Japan. It consists of custom-made baseband processing units \cite{Baseband} and commercial phased array antennas with beamforming transceivers, i.e. EVK02001 for 24 GHz and EVK06002 for 60 GHz, Sivers IMA. For each transceiver, a narrow beam in the azimuth plane is synthesized using a $2\times8$ planar array for the 24 GHz and a 16-element uniform linear array for the 60 GHz band, respectively, at both the transmitter (Tx) and receiver (Rx) side. The maximum transmission power is approximately 32 dBm for the 24 GHz and 41 dBm for the 60 GHz in terms of the equivalent isotropic radiated power (EIRP). The half power beam widths (HPBWs) of the broadside beam patterns are approximately 15$^{\circ}$ and 6$^{\circ}$ in the azimuth plane for the 24 and 60 GHz, respectively. Moreover, the HPBWs in the elevation plane of the 24 GHz phased array are 45$^{\circ}$ for both Tx and Rx, while these of the 60 GHz phased array are 18$^{\circ}$ for the Rx and 45$^{\circ}$ for the Tx. 

In our setup in Fig.~\ref{fig:sounderimage}, we have 5 Tx and 5 Rx beams covering 90$^{\circ}$ scanning range at link ends at 24 GHz. At 60 GHz, we have 11 Tx and 12 Rx beams covering 90$^{\circ}$ scanning range at link ends. The 24 GHz RF frontends were placed on top of the 60 GHz RF frontends as demonstrated in Fig.~\ref{fig:sounderimage}. 
The channel sounding system in this setup employees a dual frequency time division multiplexing (TDM) scheme to measure one 24 GHz channel and one 60 GHz channel at one time. A complete measurement consists of 132 MIMO measurement blocks for angular scanning with 11 Tx and 12 Rx beams for both bands, which takes about 5 minutes \cite{Sounder}.
The sounder utilizes multitone signals, and the multitone allocation and waveform can be found in \cite{Sounder}. The number of samples $N_f$ for a single waveform is 2048 for both frequencies. The numbers of tones are $512$ in 200 MHz bandwidth (delay resolution 5 ns) for the 24 GHz and $1024$ in 400 MHz bandwidth (delay resolution 2.5 ns) for the 60 GHz band. The maximum measurable delay is 2.56 $\mu s$, and sampling rate is 800 Msps, for both bands.

Details on the dynamic range of the system, the time synchronization and reference clock signal generation, the calibration and the transmission and reception of the measurement block of snapshots, as well as the phase/frequency offset and the trigger pulse offset can be found in \cite{Sounder}. 
Note that we do not require a strict phase coherence among multiple snapshots obtained within a measurement, because the angle-resolved power spectra are used in post-processing. The inter-measurement delay-time deviation generated during the measurement campaign can be fixed simply by using the line-of-sight (LOS) component \cite{Sounder}.

\begin{figure}[tb]
\centering
\includegraphics[width=0.7\columnwidth]{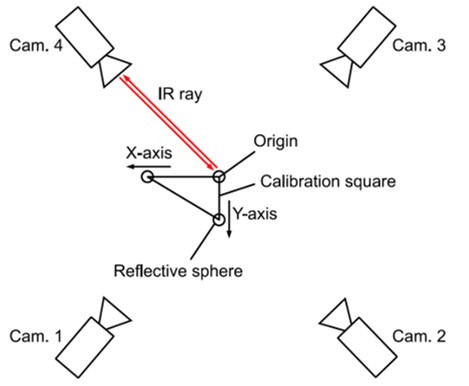}
\caption{System configuration of distributed RGBD cameras}
\label{fig:cameraimage}
\end{figure}

\subsection{The Distributed RGB-Depth Camera Network}

\begin{figure*}[tb]
\centering
\includegraphics[width=0.8\textwidth]{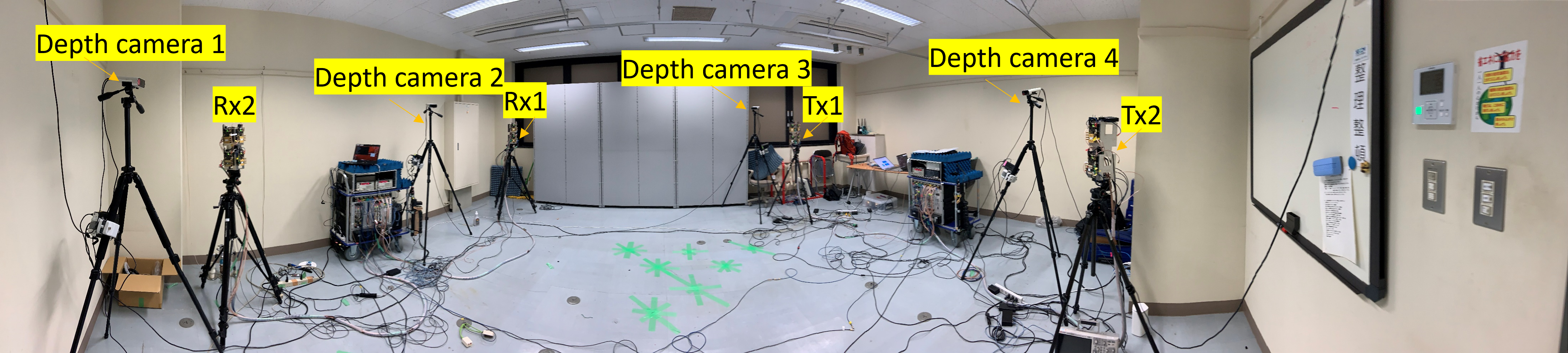}
\caption{Panoramic view of the measurement setup and the room}
\label{fig:eucap2}
\end{figure*}

As is shown in Fig.~\ref{fig:cameraimage}, the distributed RGBD camera system is composed of four Microsoft Azure Kinect DK cameras that are synchronized to an external trigger signal \cite{Kang_paper}. When the rising edge of an external trigger signal is detected by the cameras, the cameras start to record. The recorded four RGBD videos are then processed and converted to a complete point cloud by registering the local point clouds to the global coordinate system. 
During our measurement, as shown in Fig.~\ref{fig:eucap2}, four RGBD cameras are distributed in the vicinity of the TRxs of the channel sounder to capture the ground truth of the measurement (especially the human body and the relative locations to the transceivers).

The processing procedure is as follows.
First, the surface of the human body in the test area is measured independently as a point cloud in the local coordinate system of each camera. 
Second, to reconstruct a complete surface of the human body in a global coordinate, we process the global registration of the local point cloud using a designed calibration square with three infrared (IR) reflecting spheres located at the corners \cite{OptiTrack}, as is shown in Fig.~\ref{fig:cameraimage}. Using the calibration square requires us to perform a calibration measurement with it beforehand and find the transformation matrices between global and local coordinates. 
The matrices are the ones transforming the centers of the three reflecting spheres found from the IR photos to the dimension of the calibration square. 
The calibration matrices are then used to register the independent point cloud from each camera into the global coordinate system whose origin and axes are determined by the calibration square as in Fig.~\ref{fig:cameraimage}.

\subsection{The Measurement Campaign and Scenarios}

\begin{figure}[tb]
\centering
\includegraphics[width=\columnwidth]{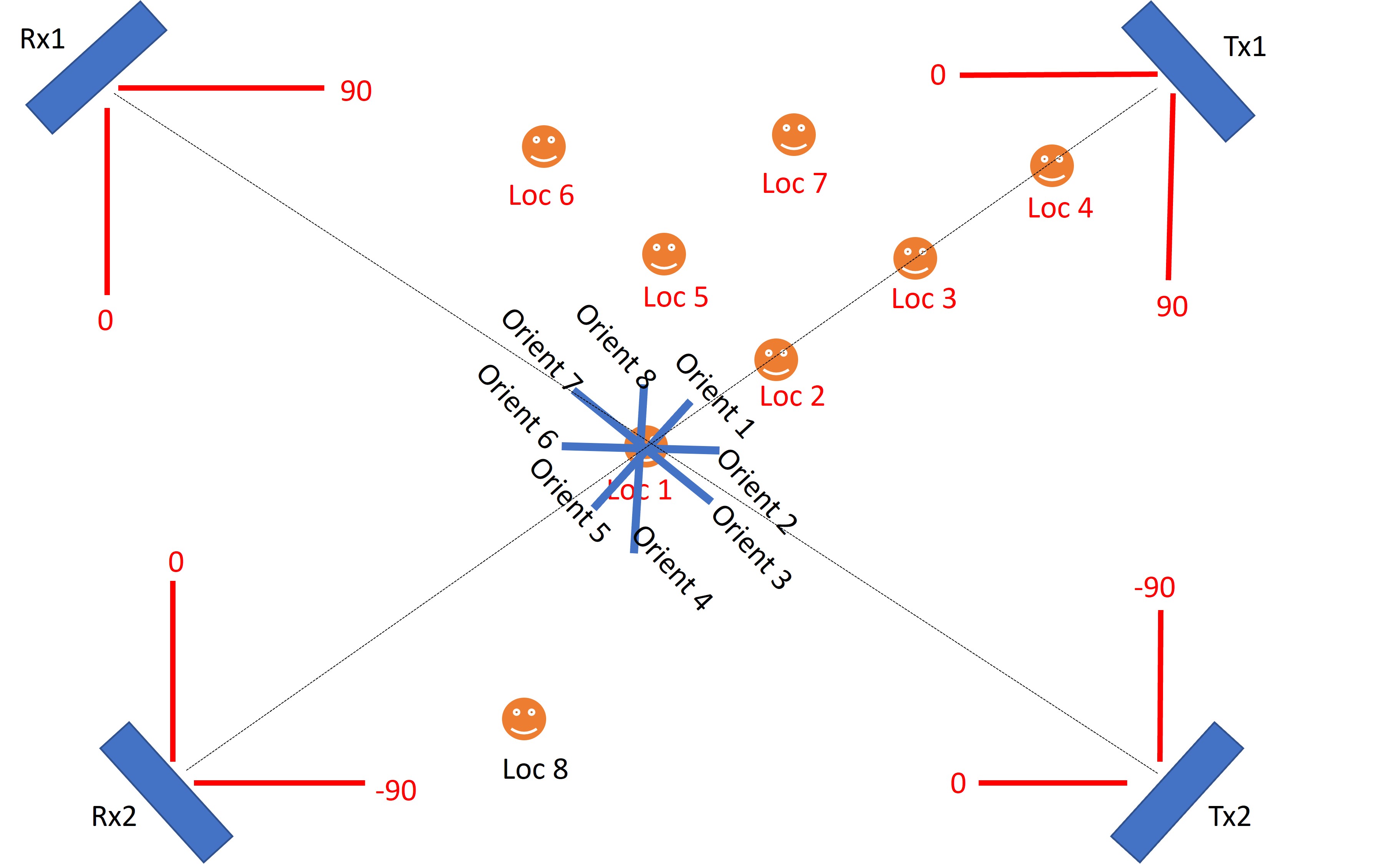}
\caption{A sketch of the measurement scenario, human body locations and orientations}
\label{fig:eucap}
\end{figure}

\begin{table}[tb]
\renewcommand{\arraystretch}{1}
\caption{Measurement scenarios with person A, B, C, D, E }
\label{table_example}
\centering
\begin{tabularx}{\columnwidth}{l|l}
\hline
Nr. person & Location $\&$ Orientation; e.g., A12 = 'Name' Loc Orient \\
\hline
1 &  A stands at from Loc1 to Loc8, each with orientation   \\
(64 meas.) & from Orient1 to Orient8 \\
\hline
2  &  B21$\_$C41, B23$\_$C43, B25$\_$C33, B26$\_$C37, B26$\_$C44, \\
(56 meas.)  &  B28$\_$C38, B32$\_$C24, B33$\_$C28, B38$\_$C26, B41$\_$C24, \\
  &   B43$\_$C27, B44$\_$C27,   \\
  &  A21$\_$C68, A21$\_$C71, A22$\_$C51, A22$\_$C71, A23$\_$C64, \\
  &  A23$\_$C72, A23$\_$C73, A23$\_$C76, A25$\_$C54, A26$\_$C56, \\
  &  A26$\_$C61, A27$\_$C73, A51$\_$C22, A51$\_$C24, A51$\_$C61, \\
  &  A51$\_$C71, A52$\_$C68, A52$\_$C78, A53$\_$C68, A53$\_$C72, \\
  &  A53$\_$C73, A54$\_$C62, A54$\_$C63, A54$\_$C73, A55$\_$C66, \\
  &  A55$\_$C75, A56$\_$C24, A56$\_$C78, A61$\_$C23, A63$\_$C21, \\
  &  A68$\_$C26, A81$\_$C21, A81$\_$C24, A81$\_$C71, A82$\_$C28, \\
  &  A82$\_$C72, A83$\_$C77, A85$\_$C27, A86$\_$C74, A87$\_$C23, \\
  &  A87$\_$C73, A88$\_$C26, A88$\_$C72, A84$\_$C77 \\
\hline
3 &  A22$\_$D76$\_$E88, A23$\_$D67$\_$E58, A23$\_$D78$\_$E81, \\
(30 meas.)    & A24$\_$D61$\_$E52, A51$\_$D24$\_$E66, A52$\_$D23$\_$E62, \\
    & A52$\_$D68$\_$E88, A53$\_$D68$\_$E71, A53$\_$D85$\_$E72, \\
    & A54$\_$D66$\_$E78, A55$\_$D61$\_$E82, A56$\_$D82$\_$E75, \\
    & A61$\_$D52$\_$E23, A61$\_$D82$\_$E56, A63$\_$D53$\_$E73, \\
    & A66$\_$D58$\_$E27, A67$\_$D55$\_$E77, A68$\_$D86$\_$E52, \\
    & A71$\_$D66$\_$E58, A71$\_$D87$\_$E26, A72$\_$D54$\_$E81, \\
    & A73$\_$D56$\_$E86, A73$\_$D68$\_$E54, A73$\_$D68$\_$E22, \\
    & A81$\_$D67$\_$E52, A81$\_$D71$\_$E52, A82$\_$D22$\_$E72, \\
    & A86$\_$D66$\_$E54, A86$\_$D77$\_$E54, A87$\_$D28$\_$E74 \\
\hline
\end{tabularx}
\end{table}

The measurements were conducted in an office room on the third floor of the General Research Building of Information Science and Technology at the Niigata University, Japan.
As is shown in Fig.~\ref{fig:eucap2}, in addition to 2 sites of Tx, 2 sites of Rx and 4 sites of depth camera, there is also a big metal cabinet in front of the windows, and on the other side a white board with polyester laminate coated steel hanging on the same wall as that of the metal door. As is shown in Fig.~\ref{fig:eucap}, the Tx1 and Rx2 array surfaces are facing to each other, while the Tx2 and Rx1 array surfaces are facing to each other. Tx1 and Rx1 are closer to the window side, while Tx2 and Rx2 are close to the door and whiteboard side (the white board is closer to Tx2). 

Moreover, we have 8 representative locations marked in the test zone, denoted by 'Loc $\#$'. Loc1, Loc2, Loc3 and Loc4 are in the direct path between Tx1 and Rx2; Loc5, Loc6, Loc7 and Loc8 are deviated from the direct line between the possible LOS links. As in Fig.~\ref{fig:eucap}, for each location, a person can stand there facing different directions, or in other words having different body orientations, denoted by 'Orient $\#$'. E.g., Orient1 is facing to the Tx1 direction, Orient2 is to face about 45 degrees clockwise from Orient1. All the measured scenarios can be found in Table~\ref{table_example}. In total 150 measurements were conducted, which is equivalent to about 13 hours measurement time excluding the preparation time.

\begin{figure*}[tb]
\centering
{\includegraphics[width=0.47\textwidth]{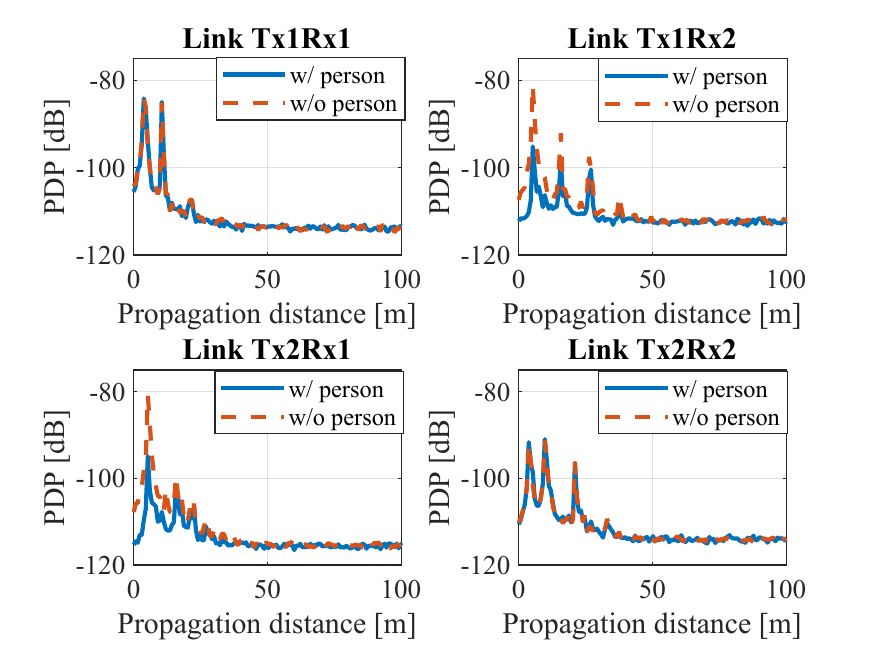}}
{\includegraphics[width=0.47\textwidth]{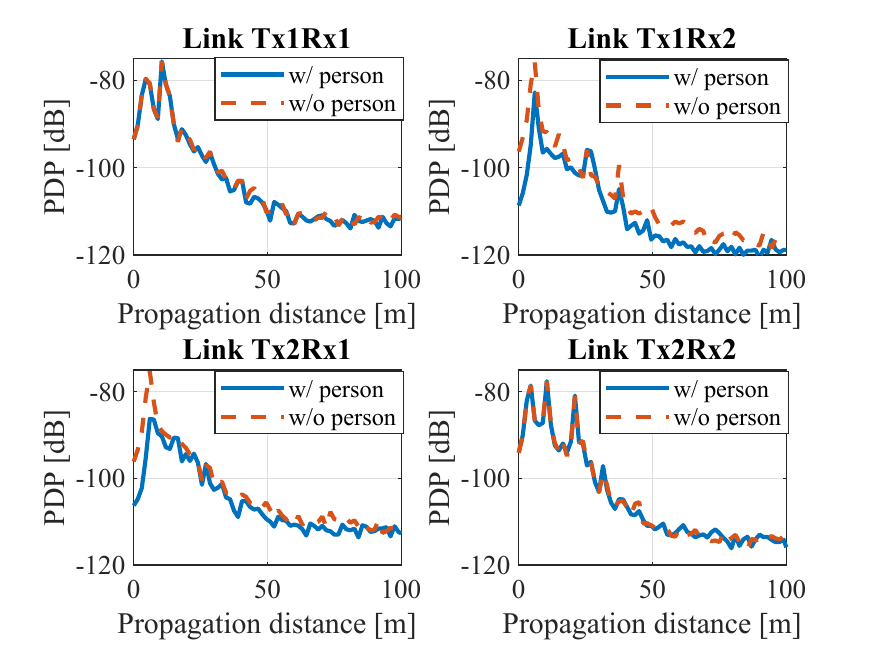}}
\caption{PDP of channel w/ and w/o person standing at Loc1 with facing orientation Orient1. \textbf{Left}: 60 GHz. \textbf{Right}: 24 GHz.}
\label{fig1}
\end{figure*}

\section{ISAC Channel Characterization with the Presence of One Person}

Due to the limited space, in this paper, we focus on analyzing the channels with the presence of 1 person with controlled locations and body facing orientations. Analysis with multiple persons will be presented in our other paper.

The measured channel transfer function (CTF) for each link and each frequency band can be denoted as 
$\bm{H} \in \mathbb{C}^{N_f, N_{\phi_R}, N_{\phi_T}}$,
where $N_{f}$ is the total number of the sampling frequency, $N_{\phi_R}$ and $N_{\phi_T}$ are the total number of the Rx beam directions and that of the Tx beam directions in the azimuth plane, respectively. We have $2\times 2$ distributed links, i.e., Tx1Rx1, Tx1Rx2, Tx2Rx1, Tx2Rx2, and dual-band, hence 8 such channel matrices $\bm{H}$ for one measurement. By taking the inverse Fourier transform to convert from the frequency $f$ domain to the delay $\tau$ domain, the CIR matrix, 
$\bm{h} \in \mathbb{C}^{N_{\tau}, N_{\phi_R}, N_{\phi_T}}$,
of each link for each frequency band can be obtained.
Its entry is denoted as ${h}_{n_{\tau}, n_{\phi_R}, n_{\phi_T}}$, where
$n_{\tau}$ is the index of delay bin, $n_{\phi_R}$ and $n_{\phi_T}$ are the indexes of the Rx and Tx beam directions in azimuth domain respectively.

The power delay profile (PDP) for each link and each frequency is then synthesized as:
\begin{equation}
\textrm{PDP}_{n_{\tau}} = \sum_{n_{\phi_R}=1}^{N_{\phi_R}} \sum_{n_{\phi_T}=1}^{N_{\phi_T}}  \left| {h}_{n_\tau, n_{\phi_R}, n_{\phi_T}}  \right|^2.
\end{equation}
The root mean square (RMS) delay spread (DS) can be further calculated as:
\begin{equation}
\sigma_{\mathrm{DS}} = \sqrt{\frac{\sum_{n_{\tau}=1}^{N_{\tau}} \tau^2 \mathrm{PDP}_{n_{\tau}}}   {\sum_{n_{\tau}=1}^{N_{\tau}} \mathrm{PDP}_{n_{\tau}}} - \left( \frac{\sum_{n_{\tau}=1}^{N_{\tau}} \tau \mathrm{PDP}_{n_{\tau}}}{\sum_{n_{\tau}=1}^{N_{\tau}} \mathrm{PDP}_{n_{\tau}}} \right)^2 }.
\end{equation} 

The angular-delay power spectrum (ADPS) at Tx and Rx sides are calculated by a summation over certain angular ranges for each link and each frequency:
\begin{equation}
\begin{split}
\textrm{ADPS}_{n_{\tau}, n_{\phi_{T}}} = \sum_{n_{\phi_{R}}=1}^{N_{\phi_{R}}} \left| {h}_{n_\tau, n_{\phi_{R}}, n_{\phi_{T}}} \right|^2, \\
\textrm{ADPS}_{n_{\tau}, n_{\phi_{R}}} = \sum_{n_{\phi_{T}}=1}^{N_{\phi_{T}}} \left| {h}_{n_\tau, n_{\phi_{R}}, n_{\phi_{T}}} \right|^2.
\end{split}
\end{equation}
The azimuth spread of departure (ASD) or azimuth spread of arrival (ASA) is calculated as:
\begin{equation}
\sigma_{\mathrm{AS}} = \sqrt{ -2 \ln \left| \frac{\sum_{n_{\Omega}} \exp{ \left(j \cdot \Omega \cdot \mathrm{PAP}_{\Omega} \right)}}{\sum_{n_{\Omega}} \mathrm{PAP}_{\Omega} } \right| },
\end{equation}
where $\Omega$ can be either $\phi_R$ or $\phi_T$ the azimuth angles of the beam directions at receiver or transmitter respectively, and $\mathrm{PAP}_{\Omega}$ is the single-direction power angular profile (PAP):
\begin{equation}
\begin{split}
\mathrm{PAP}_{\phi_{R}} &= \sum_{  n_{\phi_T}   } \sum_{   n_{\tau}  } \left| {h}_{n_\tau, n_{\phi_{R}}, n_{\phi_{T}}} \right|^2, \\
\mathrm{PAP}_{\phi_{T}} &= \sum_{  n_{\phi_R}   } \sum_{   n_{\tau}  } \left| {h}_{n_\tau, n_{\phi_{R}}, n_{\phi_{T}}} \right|^2. 
\end{split}
\end{equation}

\begin{figure}[thb]
\centerline{\includegraphics[width=0.9\columnwidth]{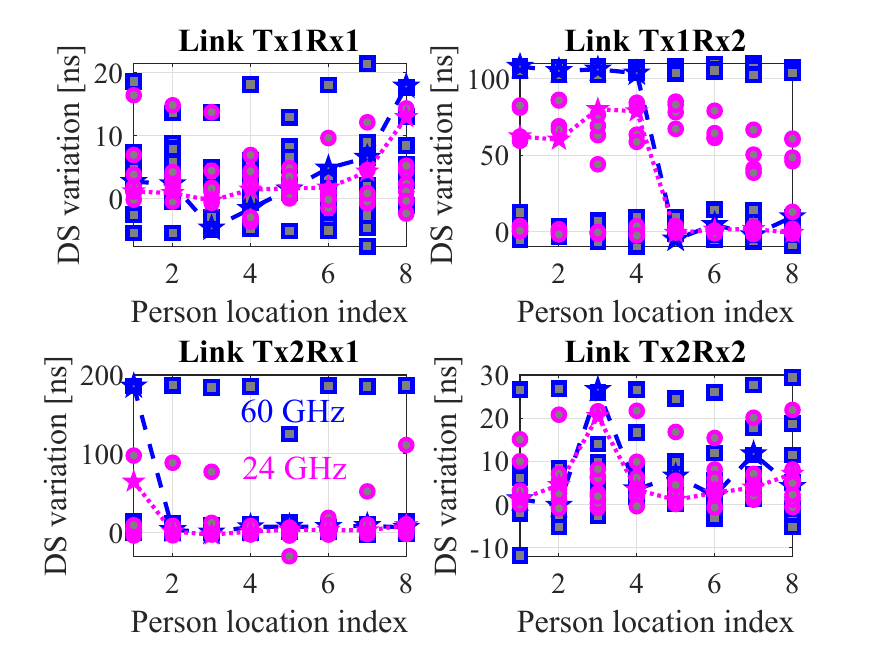}}
\caption{DS variation between w/ $\&$  w/o person for all locations and orientations; blue box values are for 60 GHz and magenta circles for 24 GHz; blue pentagram is the median among all orientations at 1 location for 60 GHz and magenta for 24 GHz. }
\label{fig2}
\end{figure}

\paragraph{Delay Dispersion}
In Fig.~\ref{fig1}, examples of PDP of the distributed links at dual bands are shown for one scenario with person A standing at Loc1 with Orient1. 
\textit{First}, the four links, when without any person, present distinct power delay profiles: Tx1Rx2 and Tx2Rx1 are with clear LOS path, at 5.3 m at 60 GHz and 6 m at 24 GHz; Tx1Rx2 contains stronger multipath in later delay bins (= larger propagation distances) compared to Tx2Rx1. Tx1Rx1 and Tx2Rx2 show typical non-LOS (NLOS) or obstructed-LOS (OLOS) channel traits with comparably strong or stronger path at the second cluster peak compared to the first peak; in Tx2Rx2, a strong peak is observed even at 10 m later (relative to the second strong peak). 
\textit{Second}, the person influences the links differently. Due to the location of the person, being residing inside both the Fresnel zone of link Tx1Rx2 and that of Tx2Rx1, the PDP changes significantly comparing the two links when w/ and w/o the person, and such changes result mainly from human blockage (body absorption). For the link Tx1Rx1 and Tx2Rx2 that are mainly influenced by human reflection, the PDP changes are less obvious compared to the blockage effect for this specific human location. Besides, comparing the two frequency bands, 60 GHz has 400 MHz bandwidth which is 2 times more delay resolution than that at 24 GHz with 200 MHz bandwidth, the peaks are more separable at 60 GHz.

Fig.~\ref{fig2} shows the quantitative results of the DS variations for all the measured 8 locations and each with 8 different orientations. The showed values are the changes of the DS when w/ and w/o person. It can be observed that with the presence of the person, depending on his/her location, different facing orientation could lead to a profound increase in the DS compared to that w/o person. Especially for link Tx1Rx2 and Tx2Rx1, the median increase of DS among all orientations when the person stands at Loc1 reaches up to 108 ns at 60 GHz and 62 ns at 24 GHz for Loc1, while that values for Tx2Rx1 are 185 ns for 60 GHz and 65 ns for 24 GHz. 
The profound median values of DS increase mainly occurs when the person stands within the first Fresnel zone of link Tx1Rx2 or Tx2Rx1; when the person stands outside of the Fresnel zone, even though the median DS increase is low, some facing orienation could result in a large DS value increase. On the other hand, for link Tx1Rx1 or Tx2Rx2, the max. DS changes induced by the presence of the person appears when the person is at Loc8 for Tx1Rx1 and at Loc3 for Tx2Rx2, which could be explained by that the presence of the person diffused further the originally reverberate multipath passing through that location. 
It is worth noting that, when the person stands at Loc3, link Tx1Rx2 is most likely influenced by the human and link Tx2Rx1 is less likely (depending on orientation); while link Tx2Rx2 is more likely to capture the human reflection induced changes, Tx1Rx1 is less likely (depending on orientation). This provides us a hint on how to deploy the distributed links for ISAC use cases: try to have at least 1 link that is not influenced by the presence of target in the test area (hence could be used for communication), and try to have at least 1 link that is reactive towards the human presence in the test area (hence could be used for sensing). 
Moreover, it shows that the Tx1Rx2 and Tx2Rx1 links have less profound DS increase at 24 GHz compared to that at 60 GHz; note that the 24 GHz array was placed on top of the 60 GHz array in the measurement, and hence the head induced DS change could be less profound compared to the torso.

\begin{figure}[tb]
\centering
{\includegraphics[width=0.9\columnwidth]{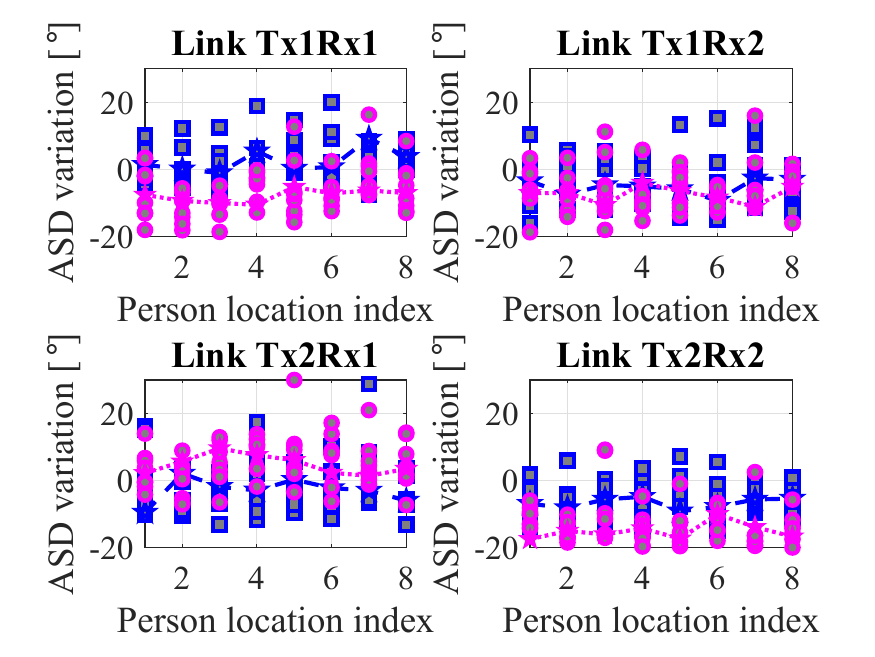}}
{\includegraphics[width=0.9\columnwidth]{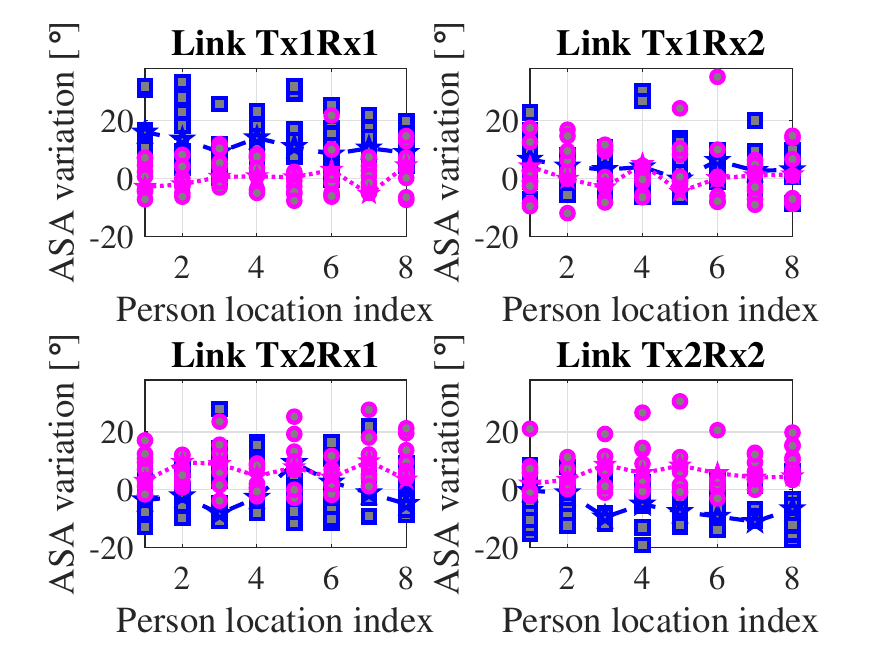}}
\caption{Top: ASD variation, Bottom: ASA variation, between the channel w/ and w/o person.}
\label{fig5}
\end{figure}

\begin{figure*}[thb]
\subfigure[60 GHz: top row is azimuth angle (Az) of departure delay power spectrum (DPS); bottom is Az of arrival DPS.]{\centerline{\includegraphics[width=0.91\textwidth]{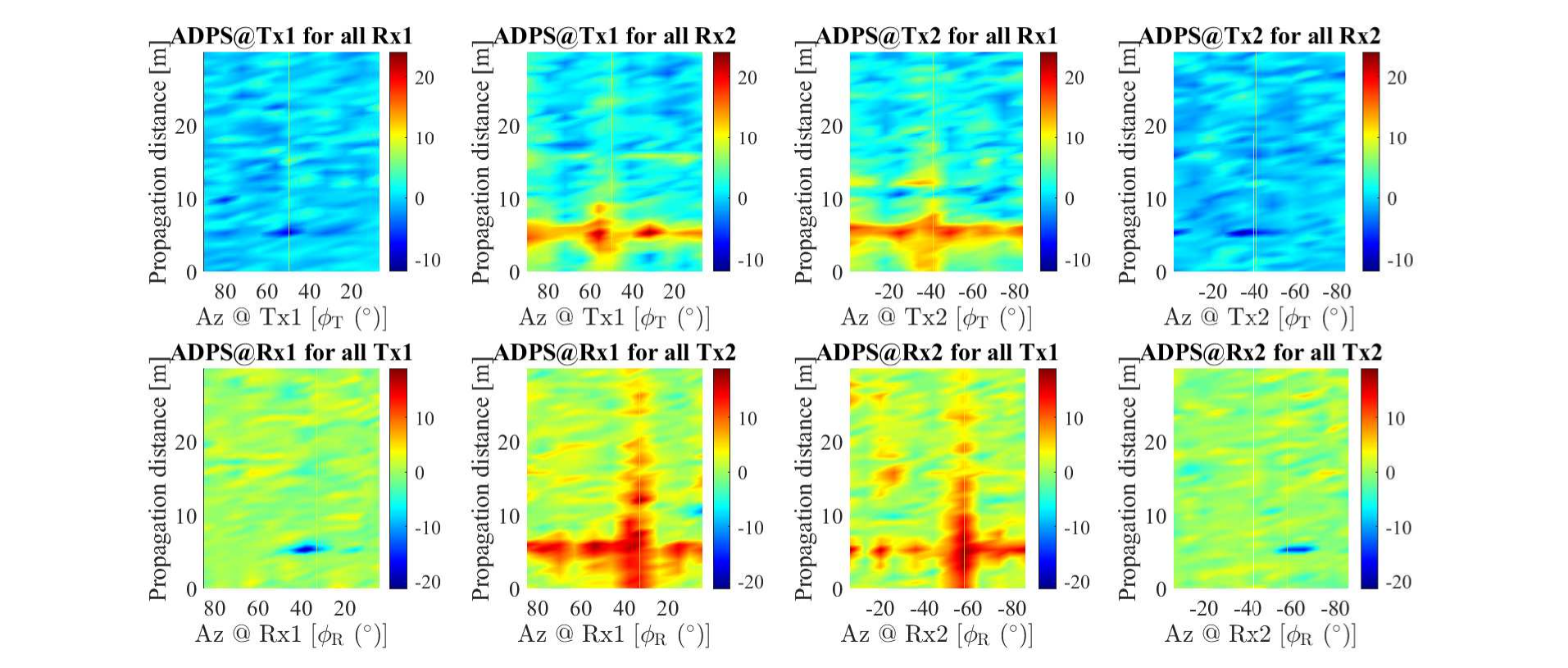}}}
\subfigure[24 GHz: top row is azimuth angle (Az) of departure delay power spectrum (DPS); bottom is Az of arrival DPS.]{\centerline{\includegraphics[width=0.91\textwidth]{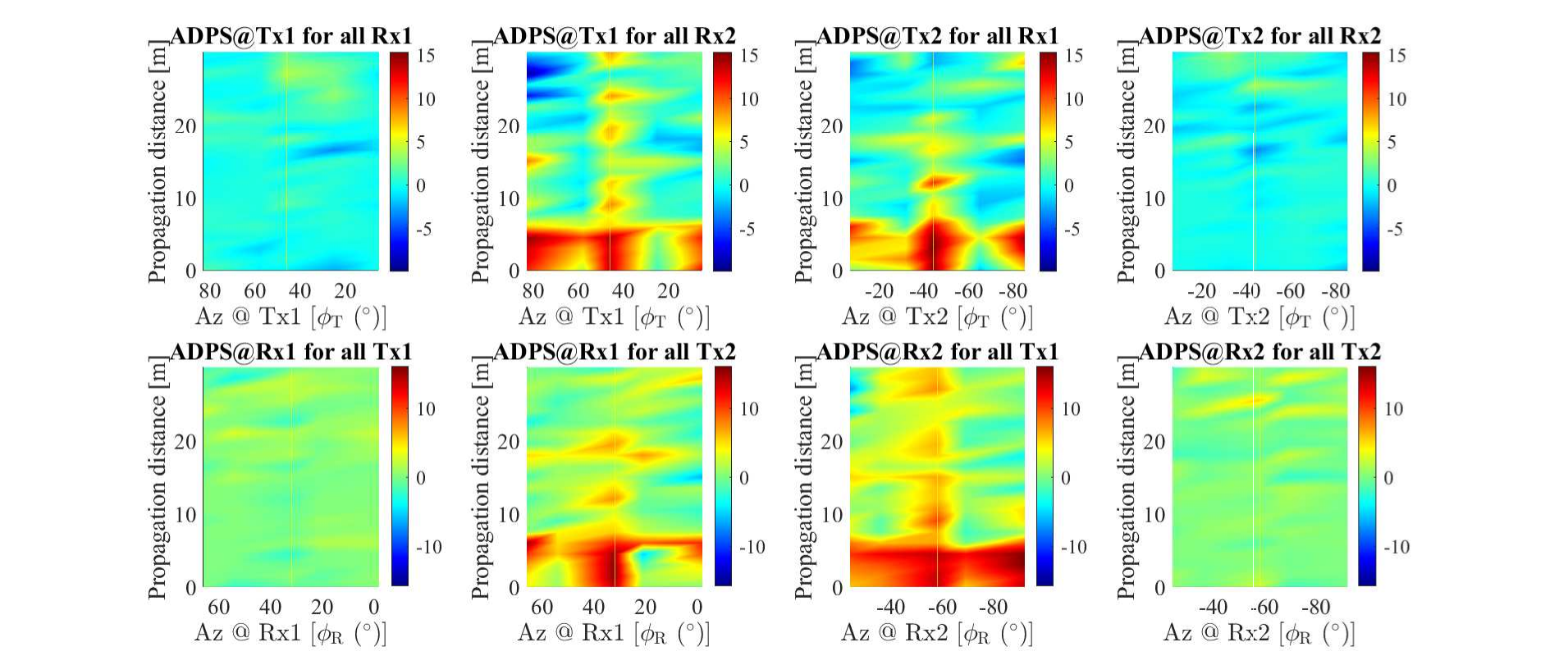}}}
\caption{ADPS changes compared to the channel w/ and that w/o the person at Loc1 with Orient1.}
\label{fig4}
\end{figure*}

   



\paragraph{Angular Dispersion}


 
Fig.~\ref{fig5} shows the quantitative results of the ASD and ASA for all the measured 8 locations and each with 8 different facing orientations. The showed values are the changes of channels w/ and w/o person. It can be observed that the person's presence could influence greatly the ASD and ASA values of the indoor channel, introducing up to 40 degrees ASA changes at 60 GHz and 33 degrees at 24 GHz, and up to 30 degrees of ASD changes for both bands. The ASA and ASD changes are influenced by the person's facing orientation significantly, causing obvious angular spread changes for all links. Interestingly, for link Tx2Rx2, the presence of the person leads to a decrease in ASD for both bands, while it leads to an increase in ASA at 24 GHz and a decrease in ASA at 60 GHz. The angular spread changes are influenced by the beamwidth and beam separation angles during beamforming, as well as the slight height difference between the heights of the transceivers of the two bands. 
Fig.~\ref{fig4} shows examples of ADPS, where the propagation distance is truncated at 30 meters in the figures. Positive values indicate power increase due to the person, while negative values indicate power decrease due to the person. It is obvious that the person does not only induce power changes at the location where he/she stands at. Interestingly, at 60 GHz, for the link Tx1Rx1 and Tx2Rx2, we can observe a power dip in the direction of the person's azimuth angle of arrival; however the power dip distance is about 5 m away from the receiver (slightly longer than the actual human distance), which could be explained by that the presence of the person caused a death of some background multipath cluster. At 60 GHz, at smaller propagation distances than 5 m, we could observe in link Tx2Rx1 and Tx1Rx2 that the power changes are mainly from the human direction.  

\section{Conclusion}
From the measurement-based channel characterization, the presence of human body in multipath rich indoor cavity could lead to profound changes in both the delay- and angular-domain signal profiles. Analysis of channels with distributed links provide insights into the deployment of link topology for ISAC use cases. The use of dual-band with different bandwidth and beamwidth results in distinct angular domain feature and in future we will look into whether we can use dual-band for complementary feature estimation in ISAC.


\begin{thebibliography}{1}

\bibitem{5GNRmmWave}
''TS 38.101-2: NR; User Equipment (UE) radio transmission and reception; Part 2: Range 2 Standalone'' (18.2.0 ed.). 3GPP. 2023-06-30.


\bibitem{WiGig}
''Status of Project IEEE 802.11ay''. Institute of Electrical and Electronics Engineers. Retrieved 21 September 2020.
  

\bibitem{CoDesign1}
H. Alidoustaghdam, et al., ''Sparse Tiled Planar Array: The Shared Multibeam Aperture for Millimeter-Wave Joint Communication and Sensing,'' \emph{Electronics} 2023, 12, 3115. 


\bibitem{CoDesign2}
C. Li, et al., ''Contact-Free Multitarget Tracking Using Distributed Massive MIMO-OFDM Communication System: Prototype and Analysis,'' in \emph{IEEE Internet of Things Journal}, vol.10, no.10, pp.9220-9233, May, 2023.

\bibitem{reverberation}
Y. Miao, et al., "Reverberant Room-to-Room Radio Channel Prediction by Using Rays and Graphs," in IEEE Transactions on Antennas and Propagation, vol. 67, no. 1, pp. 484-494, Jan. 2019.


\bibitem{Blockage4}
Y. Oguma, et al., ''Proactive Handover Based on Human Blockage Prediction Using RGB-D Cameras for mmWave Communications'', \emph{IEICE Transactions on Communications}, vol.E99.B, no.8, pp.1734-1744, 2016.

\bibitem{SKINReflectance}
T. Wu, et al., ''The human body and millimeter-wave wireless communication systems: Interactions and implications,'' \emph{2015 IEEE International Conference on Communications (ICC)}, London, UK, 2015.

\bibitem{Sounder}
M. Kim, et al., ''A 24/60-GHz Dual-Band Double-Directional Channel Sounder Using COTS Phased Arrays,'' \emph{2022 IEEE International Conference on Communications Workshops}, Seoul, Korea, 2022.

\bibitem{Baseband}
M. Kim, et al., ''Fast Double-Directional Full Azimuth Sweep Channel Sounder Using Low-Cost COTS Beamforming RF Transceivers,'' in \emph{IEEE Access}, vol. 9, pp. 80288-80299, 2021.

\bibitem{Kang_paper}
C. Kang, et al., "Synchronized Dynamic Channel Sounder and Posture Capture for Millimeter Wave Radio Channel Suffered from Human Body Shadowing," 17th EuCAP, Florence, Italy, 2023.

\bibitem{OptiTrack}
Calibration Squares, https://docs.optitrack.com/motive/calibration/cali-\\
bration-squares.



\end{thebibliography}
\end{document}